\documentclass[apl,preprint,showpacs]{revtex4}
\input epsf
\usepackage{amsmath}
\usepackage{amssymb}
\begin{document}

\title{Robust coupling of superconducting order parameter in a mesoscale NbN-Fe-NbN epitaxial structure}

\author{S. K. Bose and R. C. Budhani}

\email{rcb@iitk.ac.in}

\affiliation{Condensed Matter - Low Dimensional Systems
Laboratory, Department of Physics, Indian Institute of Technology
Kanpur, Kanpur - 208016, India}

\date{\today}

\begin{abstract}

We report an unconventional and promising route to self-assemble
distributed superconductor-ferromagnet-superconductor (S-F-S)
Josephson Junctions on single crystal [100] MgO. These structures
consist of [110] epitaxial nano-plaquettes of Fe covered with
superconducting NbN films of varying thickness. The S-F-S
structures are characterized by a strong magnetoresistance (MR)
anisotropy for the in-plane and out-of-plane magnetic fields. The
stronger in-plane MR suggests decoherence of S-F-S junctions whose
critical current follows a ($1-\emph{T/T$_c$}$) and
($1-\emph{T/T$_c$}$)\emph{$^{1/2}$} dependence for \emph{T}
$\approx$ \emph{T$_c$} and \emph{T} $\ll$ \emph{T$_c$}
respectively, in accordance with the theory of supercurrent
transport in such junctions.

\end{abstract}


\maketitle

The tunneling of Cooper pair order parameter through a thin
barrier separating two bulk superconductors, as predicted by
Josephson\cite{josephson} and seen subsequently in thin film
junctions\cite{anderson} has had a far reaching impact on physics
and technology. The applications of Josephson Junctions (JJs)
range from sensors for ultralow magnetic fields and weak
electromagnetic radiation\cite{krey}, millimeter wave
resonators\cite{klushin}, programmable voltage
standards\cite{schulze}, superconducting flux
qubits\cite{mooji_chi}, etc. The physics of JJs changes remarkably
when the barrier material is a ferromagnet\cite{buzdin,ryazanov}.
Interesting effects with rich underlying physics and many
promising applications are expected in superconductor-ferromagnet
(S-F) hybrids of nanoscale dimensions\cite{makhlin}. In recent
years nanoscale S-F structures have been synthesized using the
conventional approaches of nano-lithography such as electron beam
patterning, atomic force microscopy, and focused ion beam milling.

In this letter we report observation of a giant anisotropic
magnetoresistance (MR) in self-assembled
nanometer-scale-distributed junctions of Fe and superconducting
NbN. Our methodology of synthesis utilizes stress-tuned
Volmer-Weber (VW) type\cite{jordan} plaquette growth of Fe on
[100] MgO, whose electrical connectivity is tuned by NbN layers of
different thickness (\emph{d$_{NbN}$}) deposited on top of the VW
template. A KrF excimer laser (\emph{$\lambda$} $=$248 nm) based
pulsed laser ablation technique was used to deposit the
nanostructured Fe and epitaxial NbN thin films as described in our
earlier works\cite{bose,senapati}. In brief, the growth and the
post-growth annealing temperature for Fe plaquettes was $\sim$700
$^\circ$C whereas the NbN layer was deposited at 200 $^\circ$C to
inhibit the formation of iron nitride at NbN-Fe interfaces. The
nominal thickness of the Fe base layer is 40 nm, whereas
\emph{d$_{NbN}$} =10, 20, and 30 nm have been used.

\begin{figure}
\centerline{\epsfxsize = 11 cm \epsfbox{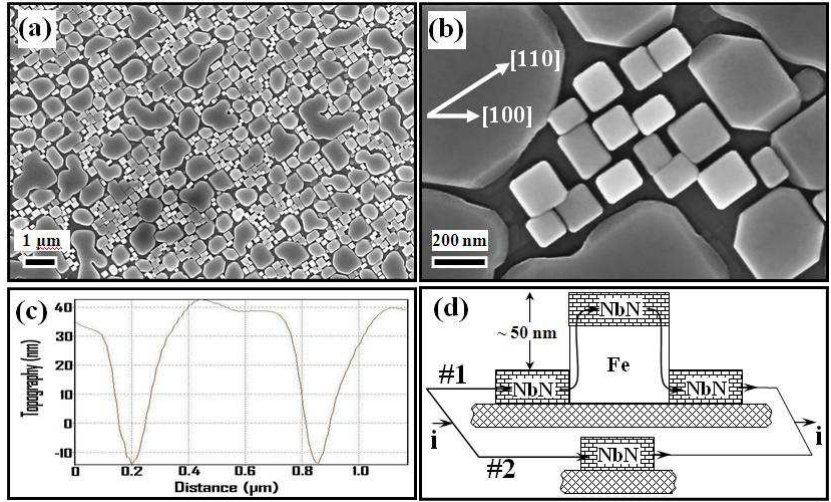}}
\caption{\label{fig1} (a \& b) Scanning electron micrographs of
the Fe(40 nm) / NbN(30 nm) sample. The top of nearly perfect
square Fe plaquettes and the inter-plaquette space is covered with
epitaxial NbN. (c) A typical atomic-force-microscope line scan
showing the height of Fe plaquettes. (d) A sketch showing the two
distinct parallel paths for the flow of supercurrent through the
S-F-S hybrid. Path $\sharp$1 is for the NbN-Fe-NbN junctions and
$\sharp$2 for current flow through the percolating backbone of
NbN.}
\end{figure}

The scanning electron micrographs (SEM) of the samples as shown in
Fig. 1(a \& b) reveal that the Fe template consists of nearly
perfect square plaquettes of $\approx$100 $\times$ 100 nm$^2$
area, separated by $\approx$20 nm gaps and are aligned along MgO
[110] direction. The NbN grows epitaxially on the plaquettes and
the inter-plaquette gaps. This has been confirmed with x-ray
diffraction and x-ray fluorescence mapping of niobium. A typical
atomic force microscope (AFM) line scan shown in Fig. 1(c) reveals
that the requirement of mass conservation makes the Fe
nano-plaquettes thicker than the programmed thickness of
$\approx$40 nm.  Due to this unique structure, the flow of
supercurrent in this S-F hybrid occurs through two parallel
channels as shown in Fig. 1(d). One of these paths (i.e.
$\sharp$1) is the double S-F-S junction, in which supercurrent
goes vertically up through thin Fe layers into the intra-plaquette
NbN and then comes down, again through the Fe. The other route
($\sharp$2) is the thin percolating backbone of NbN in the
inter-plaquette spaces. As will be shown later, the path $\sharp$1
involving double S-F-S junctions is of greater significance for
supercurrent transport in these hybrids.

\begin{figure}
\centerline{\epsfxsize = 10 cm \epsfbox{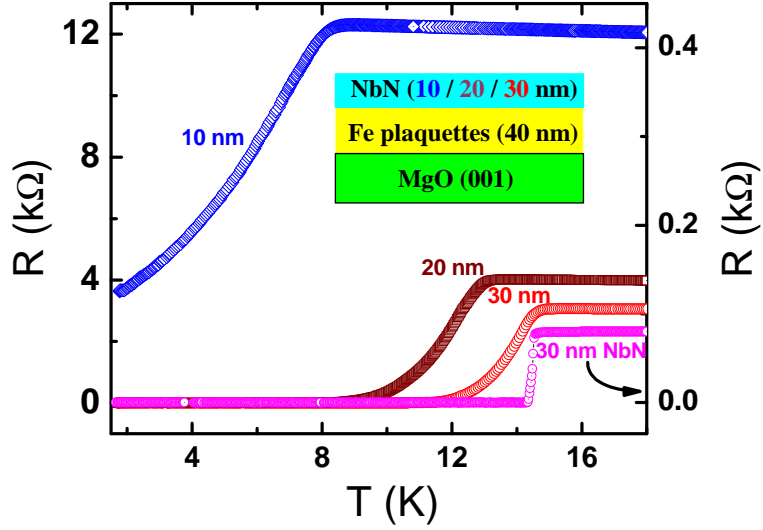}}
\caption{\label{fig2} (Color online) The temperature dependence of
resistance of the 10, 20 , 30 nm NbN covered Fe plaquettes (left
Y-scale) and 30 nm thick pure NbN film (right Y-scale). The onset
temperature of superconductivity (\emph{T$_{onset}$}) for 30 nm
hybrid is the same as of plane 30 nm NbN film, whereas the
\emph{T$_c$} and \emph{$\Delta$T$_c$} are affected significantly
when the NbN cover layer thickness is reduced.}
\end{figure}

Fig. 2 shows the superconducting (SC) transition measured across
bridges of $\approx$75 $\times$ 1300 $\mu$m$^2$ area, created by
Ar$^+$ ion milling of the S-F hybrids. For comparison the
SC-transition of a plane 30 nm thick NbN film is also shown in
Fig. 2. A significant drop in transition temperature
(\emph{T$_c$}) and increase in the width of the transition
(\emph{$\Delta$T$_c$}) along with a gain in the normal state
resistance (\emph{R$_n$}) is seen as \emph{d$_{NbN}$} is reduced
from 30 to 10 nm. In fact, the \emph{T$_c$} drops nearly twice as
fast for the hybrid ($\approx$49\%) in comparison to the drop seen
in a plane NbN film ($\approx$24\%)\cite{bose1} as the film
thickness \emph{d$_{NbN}$} is reduced from 30 to 10 nm. The
\emph{R$_n$} increases with decrease in temperature, but remains
lower than the quantum resistance for Cooper pairs (\emph{R$_Q$}
$=$ \emph{h/4e$^2$} $\sim$6.4 k$\Omega$/$\Box$), above which a
superconductor-insulator transition is seen in granular
films\cite{hebard_yazdani}. We also notice that although the
\emph{R$_n$} of the hybrid with 30 nm thick NbN is three orders of
magnitude higher compared to the \emph{R$_n$} of pure NbN film,
the onset temperature of superconductivity (\emph{T$_{onset}$}
$\approx$14.7 K) remains nearly the same in the two cases. The
\emph{$\Delta$T$_c$} ($\approx$2 K) of this hybrid, however is
seven times higher than of pure NbN ($\approx$0.3 K). This
observation suggests that superconductivity in inter-plaquette
epitaxial NbN sets in at $\approx$14.7 K, but the realization of
the zero-resistance state depends on the strength and phase factor
of the supercurrent through the double S-F junctions and the
narrow constrictions of the epitaxial NbN backbone as sketched in
Fig. 1(d).

\begin{figure}
\centerline{\epsfxsize = 10 cm \epsfbox{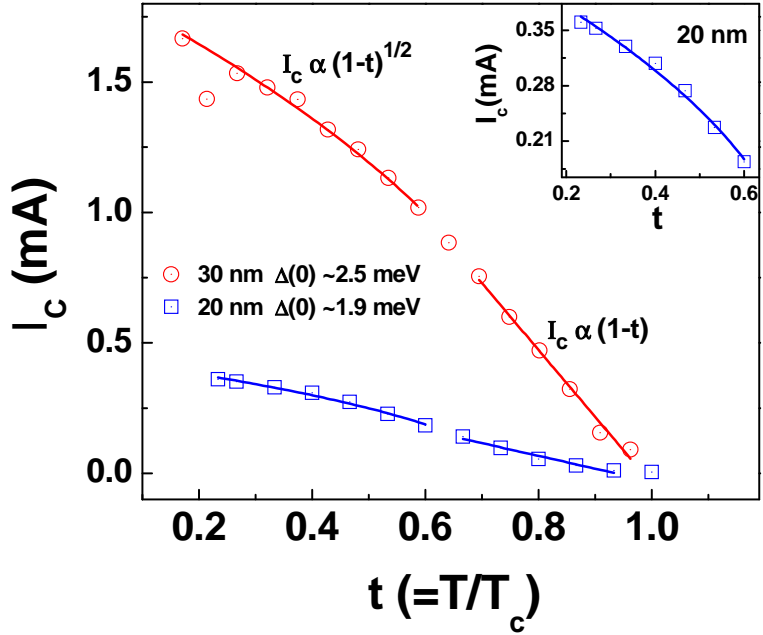}}
\caption{\label{fig3} (Color online) Critical current
(\emph{I$_c$}) variation with temperature for the 30 and 20 nm NbN
covered Fe plaquettes. Two distinct regimes of temperature
dependence can be seen in the figure. At \emph{t} $>$ 0.7 it goes
as $\sim$ (\emph{1-t}) and for \emph{t} $<$ 0.6 the dependence is
of the type (\emph{1-t})\emph{$^{1/2}$}. Inset shows the
(\emph{1-t})\emph{$^{1/2}$} dependence of the 20 nm NbN sample
over an expanded scale.}
\end{figure}

In Fig. 3 we show the critical current (\emph{I$_c$}) of two films
(\emph{d$_{NbN}$} $\approx$20 $\&$ 30 nm) in the
\emph{t}(=\emph{T/T$_c$}) range of 0.2 to 1.0. The \emph{I$_c$} of
a proximity coupled S-N-S junction depends on the number of
Andreev bound states (ABS) in the normal metal (N) spacer. If the
spacer is a ferromagnet, then the spectrum of ABS is affected by
the extent of exchange splitting of its conduction
band\cite{kontos}. It has been shown that as long as the
ferromagnet thickness \emph{d$_F$} $\ll$ \emph{$\xi$$_F$}, the
coherance length in F layer, defined as \emph{$\xi _F$}  = $\sqrt
{{\raise0.7ex\hbox{\emph{${\hbar D}$}} \!\mathord{\left/
 {\vphantom {{\hbar D} I}}\right.\kern-\nulldelimiterspace}
\!\lower0.7ex\hbox{$I$}}}$, where \emph{D} is the diffusion
coefficient and \emph{I} the exchange splitting of the
ferromagnet, the maximum current (\emph{I$_c$}) through the
junctions is given as\cite{buzdin};

\begin{equation} \label{Ic}
I_c (T) = {\rm{ }}\frac{{{\rm{32}}\sqrt {\rm{2}}
({\raise0.7ex\hbox{$\Delta $} \!\mathord{\left/
 {\vphantom {\Delta  e}}\right.\kern-\nulldelimiterspace}
\!\lower0.7ex\hbox{$e$}})}}{{R_N }}\textrm{ }{\rm{\emph{F}}}\left(
{{\Delta \mathord{\left/ {\vphantom {\Delta  T}} \right.
 \kern-\nulldelimiterspace} T}} \right)\textrm{ }{\rm{ y }}\textrm{ }{\rm{exp}}\left( {{\rm{ - y}}} \right){\rm{ sin}}\left( {{\rm{y  +  }}{\pi  \mathord{\left/
 {\vphantom {\pi  {\rm{4}}}} \right.
 \kern-\nulldelimiterspace} {\rm{4}}}} \right)
\end{equation}

where,
\[
{\rm{ y  =  }}\frac{{{\rm{\emph{d}}}_{\rm{\emph{F}}} }}{{\xi
_{\rm{F}} }}\left( {\frac{{2I}}{{\pi T_c }}} \right)^{1/2}
\]

Here, \emph{F($\Delta$/T)} has limiting values of $\frac{\pi
}{{128}}\left( {{\raise0.7ex\hbox{$\Delta $} \!\mathord{\left/
 {\vphantom {\Delta  {T_c }}}\right.\kern-\nulldelimiterspace}
\!\lower0.7ex\hbox{${T_c }$}}} \right)$ for \emph{T} $\approx$
\emph{T$_c$} and 0.071 for \emph{T} $\ll$ \emph{T$_c$}. For a
given \emph{d$_F$} and a BCS temperature dependence of the gap
parameter \emph{$\Delta$(T)} $\simeq$
3.2\emph{k$_B$T$_c$$\sqrt{1-t}$}, we get \emph{I$_c$}(\emph{T})
$\sim$ (\emph{$1-t$}) for \emph{T} $\approx$ \emph{T$_c$} and
$\sim$ \emph{$\sqrt{1-t}$} for \emph{T} $\ll$ \emph{T$_c$}. This
temperature dependence of \emph{I$_c$} is similar to that
predicted by Ambegaokar and Baratoff for weakly coupled granular
superconductors provided the suppression of gap parameter by
supercurrent is not significant\cite{clem,AB}. The data in Fig. 3
have been fitted with the (\emph{$1-t$}) and \emph{$\sqrt{1-t}$}
dependence for 1.0 $\gtrsim$ \emph{t} $\gtrsim$ 0.7 and 0.6
$\gtrsim$ \emph{t} $\gtrsim$ 0.2, respectively. This dependence of
\emph{I$_c$} (\emph{t}) fits well over the entire temperature
range and leads to \emph{$\Delta$}(0) values of 2.5 meV for 30 nm
and 1.9 meV for 20 nm case respectively, which matches well with
the BCS energy gap [2\emph{$\Delta$}(0) $=$ 3.5\emph{k$_B$T$_c$}]
values of 2.2 meV (30 nm) and 1.93 meV (20 nm). To further point
out the quality of fit, the inset shows the magnified version of
\emph{$\sqrt{1-t}$} dependence of \emph{I$_c$} for the 20 nm case.

In order to address the magnetic field (H) dependence of the
coupling between the NbN layers, we investigate the state of the
magnetization (M) of the Fe plaquettes. In Fig. 4 we show the M-H
plots taken at 5 K, with H applied along in-plane
(H$_{\parallel}$) and out-of-plane (H$_{\perp}$) direction. The
out-of-plane M-H shows that the moment of Fe plaquettes is in the
plane of the film in agreement with the earlier
studies\cite{park}.

\begin{figure}
\centerline{\epsfxsize = 10 cm \epsfbox{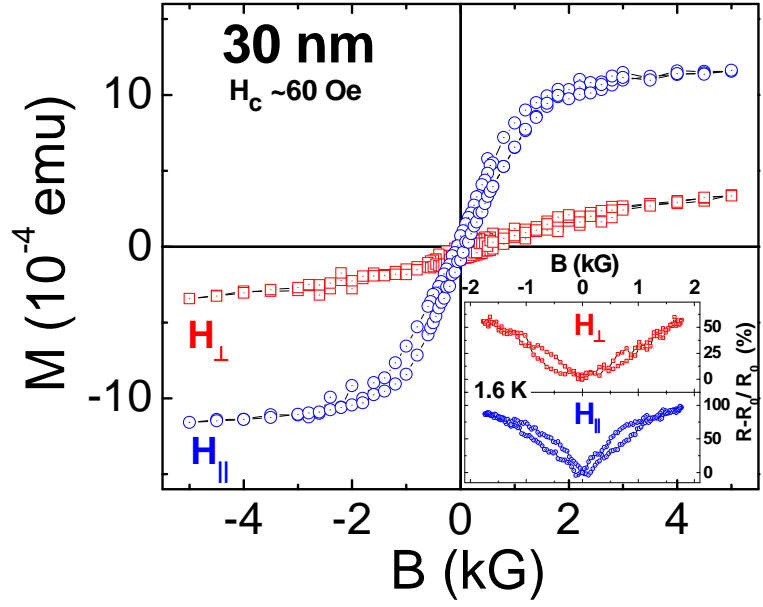}}
\caption{\label{fig4} (Color online) Magnetization of the 30 nm
NbN covered Fe plaquettes measured at 5 K with in-plane
(\emph{H}$_{\parallel}$) and out-of-plane (\emph{H}$_{\perp}$)
fields. Inset shows the magnetoresistance measured at 1.6 K. The
positive MR is twice as large for the \emph{H}$_{\parallel}$,
which is distinctly different from the behavior of a plane
superconducting film where one sees a much stronger effect of
\emph{H}$_{\perp}$.}
\end{figure}

The in-plane anisotropy of magnetization as shown in Fig. 4
affects Josephson coupling in the S-F-S junctions significantly as
seen through measurements of MR, carried out at currents exceeding
the critical current (\emph{I$_c$}). While details of the angular
dependence of MR will be presented elsewhere, in the inset of Fig.
4 we show the MR measured at 1.6 K as a function of applied field
strength. It is clear from the figure that the positive MR at a
peak field of 1500 Oe is higher by a factor of two when the field
is in the plane of the film. This is a very striking result in
view of the fact that for thin superconducting films it is the
out-of-plane field which contributes significantly to MR, due to a
copious motion of Abrikosov vortices, which is unlikely to be
arrested by a possible weak pinning by the inhomogeneous
magnetization of Fe plaquettes. A stronger response to the
H$_{\parallel}$ seen here is consistent with the fact that the
phase of the tunneling order parameter is affected significantly
when the field is in the plane of the junctions. We also note that
the MR for H$_{\parallel}$ shows two distinct cusps at $\pm$113
Oe, whose position is higher than the coercive field (\emph{H$_c$}
$\sim$60 Oe) as measured by SQUID at 5 K. This observation can be
understood in the context of number of magnetic entities actually
taking part in the magnetization reversal process. Magnetization
measurement reflects the total average response of all the
magnetic plaquettes, whereas in MR the transport current samples
only a fraction of the magnetic plaquettes which fall on its
path\cite{bose}.

In summary, we have provided a unique approach for fabrication of
distributed S-F Josephson-Junctions of nanometer length scale. Our
self-assembled NbN-Fe-NbN hybrids on [100] MgO shows
$\approx$100\% MR for \emph{H$_{\parallel}$} which is higher by a
factor of two as compared to the MR for \emph{H$_{\perp}$}. This
large in-plane response suggests breaking of phase coherence in
S-F-S junctions by the planar field. Temperature dependence of
supercurrent in these self-assembled structures  is consistent
with the theory of supercurrent transport in S-F-S junctions.

This research has been supported by a grant from the Department of
Science \& Technology under its Nanoscience \& Technology
Initiative and by the Board for Research in Nuclear Science. S. K.
Bose acknowledges financial support from the Council for
Scientific and Industrial Research, Government of India. Technical
help of P. C. Joshi, H. Pandey and R. Sharma are acknowledged.

\end{document}